\documentclass[aps,pra,reprint,superscriptaddress,floatfix]{revtex4-1}   
\pdfoutput=1         
\usepackage{hyperref}
\usepackage{graphicx}
\usepackage{amsmath}
\usepackage{upgreek}
\newcommand{\dd}{{\textrm{d}}}

\begin{document}

\title{Vorticity and quantum turbulence in the merging of superfluid Helium nanodroplets}

\author{Jos\'e Mar\'{\i}a Escart\'{\i}n}
\affiliation{Theory of Condensed Matter Group, Cavendish Laboratory, University of Cambridge,
19 JJ Thomson Avenue, Cambridge CB3 0HE, United Kingdom.
}

\author{Francesco Ancilotto}
\affiliation{Dipartimento di Fisica e Astronomia ``Galileo Galilei''
and CNISM, Universit\`a di Padova, via Marzolo 8, 35122 Padova, Italy}
\affiliation{CNR-IOM, via Bonomea, 265 - 34136 Trieste, Italy.}

\author{Manuel Barranco}

\affiliation{Departament de F\'{\i}sica Qu\`antica i Astrof\'{\i}sica, Facultat de F\'{\i}sica,
Universitat de Barcelona, Av.\ Diagonal 645,
08028 Barcelona, Spain.}
\affiliation{Institute of Nanoscience and Nanotechnology (IN2UB),
Universitat de Barcelona, Barcelona, Spain.}
\affiliation{Universit\'e Toulouse 3, Laboratoire des Collisions, Agr\'egats et
R\'eactivit\'e,
IRSAMC, 118 route de Narbonne, 31062 Toulouse Cedex 09, France.}

\author{Mart\'{\i} Pi}
\affiliation{Departament de F\'{\i}sica Qu\`antica i Astrof\'{\i}sica, Facultat de F\'{\i}sica,
Universitat de Barcelona, Av.\ Diagonal 645,
08028 Barcelona, Spain.}
\affiliation{Institute of Nanoscience and Nanotechnology (IN2UB),
Universitat de Barcelona, Barcelona, Spain.}

\begin{abstract}
We have studied the merging of two identical $^4$He droplets at zero temperature,
caused by their Van der Waals mutual attraction.
During the early stages of the merging, density structures appear
which closely match the experimental observations
by Vicente \textit{et al.} [J.\ Low Temp.\ Phys.\ \textbf{121}, 627 (2000)]. When the droplets are merging,
quantized vortex-antivortex ring pairs nucleate at the surface and
annihilate inside the merged droplet producing a roton burst.
We also observe the nucleation of quantized vortex-antivortex rings that wrap the droplet surface and
remain localized on the surface until they eventually decay into short-wavelength surface waves.
Analysis of the kinetic energy spectrum discloses the existence of a regime where
turbulence caused by vortex interaction and annihilation  is characterized by a Kolmogorov power law.
This is followed by another regime where roton radiation --produced by vortex-antivortex annihilation-- dominates,
whose hallmark is a weak, turbulent surface dynamics.
We suggest that similar processes might appear in superfluid helium droplets
after they capture impurities or if they are produced by hydrodynamic instability of a liquid jet. Experiments on collisions
between recently-discovered self-bound Bose-Einstein condensates should display a similar phenomenology.
\end{abstract}
\date{\today}

\maketitle

Superfluid helium droplets are fascinating objects that have
drawn the interest of both experimentalist and theoreticians \cite{Toe04}.
This interest is notoriously broad and includes, e.g.,
the nature of superfluidity at the nanoscale,
the interaction of atomic and molecular impurities with the hosting
droplets, and the study of vortical states in nanodroplets,
see Refs.~\onlinecite{Anc17,Mau18} for recent comprehensive reviews.

One of the hallmarks of superfluidity is the appearance
of quantized vortices \cite{Bar01,Pit16}. Whereas the presence
of vortices in macroscopic samples of bulk liquid helium was
unambiguously proved long ago, in the case of helium droplets it remained elusive.
It is only recently that vortex arrays inside droplets made of $10^8$--$10^{11}$ atoms
were detected by coherent x-ray scattering \cite{Gom14}.

The coalescence of superfluid $^4$He drops of several
tenths of cm radius levitating in a magnetic trap was investigated some time ago \cite{Vic00}.
The experimental setup allowed
to study their merging at fairly low velocity, and the
temperature was kept low enough to make the helium vapor
around them negligible, so that the droplets were essentially isolated.

In this work we address the coalescence  of two identical $^4$He  droplets drawn
together by the mutual Van der Waals (VdW) long-range attraction.
After showing that the merging  yields density structures
which closely match  the experimental observations even in their fine details,
we show that vorticity nucleates
both
inside and on the surface of the merged droplet. The subsequent decay of
vorticity results in the appearance of two distinct turbulence regimes, which we characterize with their
kinetic energy spectra. Multimedia content is provided as Supplemental Material \cite{SM}.

We study the strongly-correlated superfluid $^4$He
by using a realistic Density Functional Theory (DFT) approach
which allows to reproduce complex dynamical phenomena
such as vortex nucleation and vortex-density waves interactions \cite{Anc17}.
One important outcome of our simulations is an accurate description of
the dynamics of vortex interactions and annihilation, which is a fundamental ingredient
of current studies of quantum turbulence in liquid He and in cold-gas superfluids \cite{Vin02,Tsu09,Bar14,Tsu17}.

The Gross-Pitaevskii (GP) approach \cite{Pit16} has been also shown
to be capable to sustain these phenomena.
Vortex ring emission and the possible transition to a chaotic turbulent regime due to vortex interactions and decay
has been studied within time-dependent GP theory in both superfluid $^4$He and Bose-Einstein condensates
(BEC) \cite{Bar01,Vil18,Ber01,Fri92,Nav16,Kob05,Pro09}. It is well known, however, that GP theory
cannot provide an accurate description of the vortex  core structure in superfluid liquid helium,
and can at most reproduce the phonon part of the  $^4$He dispersion relation.
In the following,
we will prove the role of roton emission
in turbulence phenomena in $^4$He, and we will show that the roton wave-vector provides indeed
a natural dividing line between the different length scales contributing to the
turbulence created by the droplet collision process.

\begin{figure}[hbt!]
\centerline{\includegraphics[width=1.0\linewidth,clip]{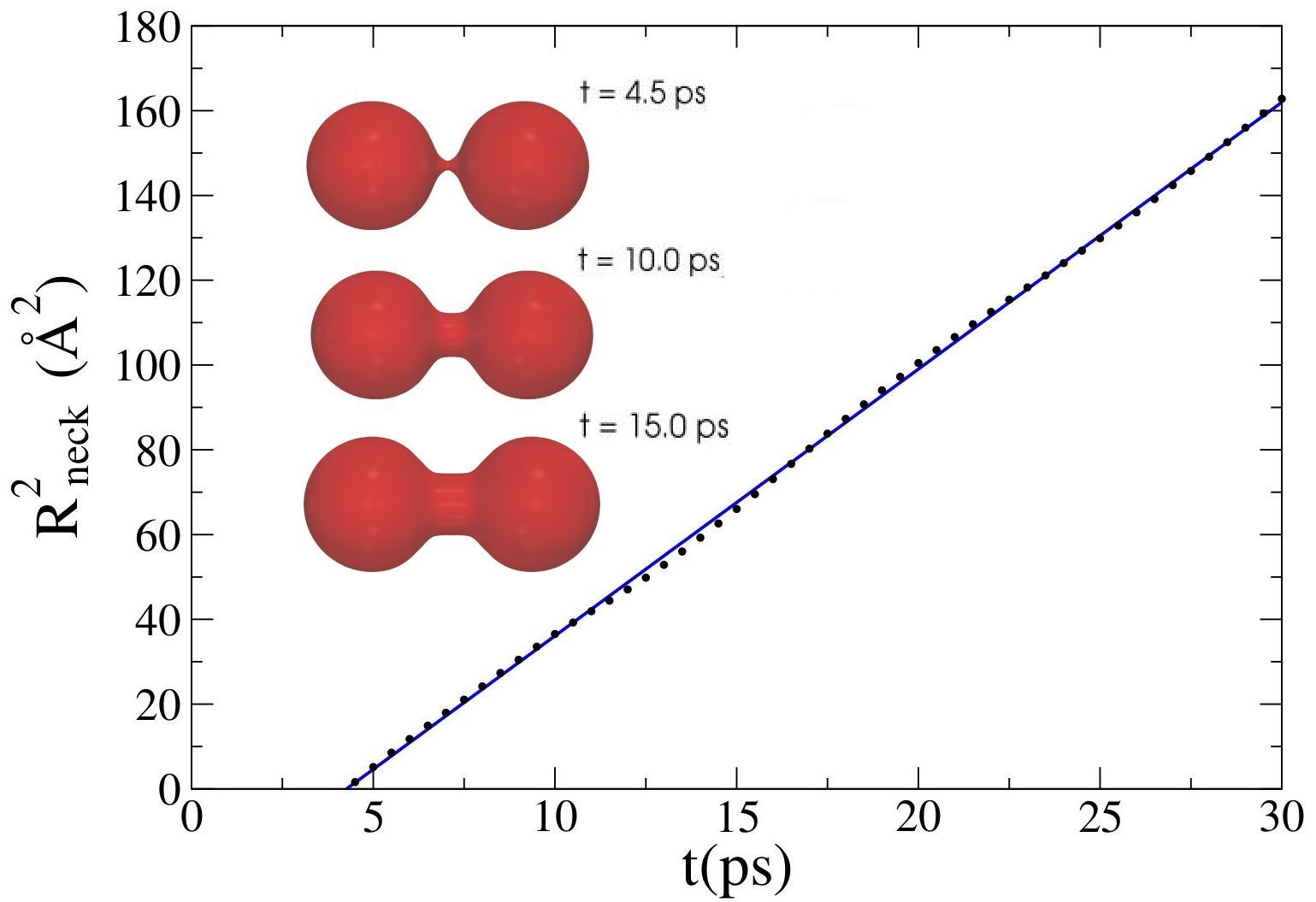}}
\caption{%
Time evolution of $R_{\textrm{neck}}^2$. Also shown are several droplet configurations
at the labeled times.}
\label{fig1}
\end{figure}

\begin{figure}[hbt!]
\centerline{\includegraphics[width=0.8\linewidth,clip]{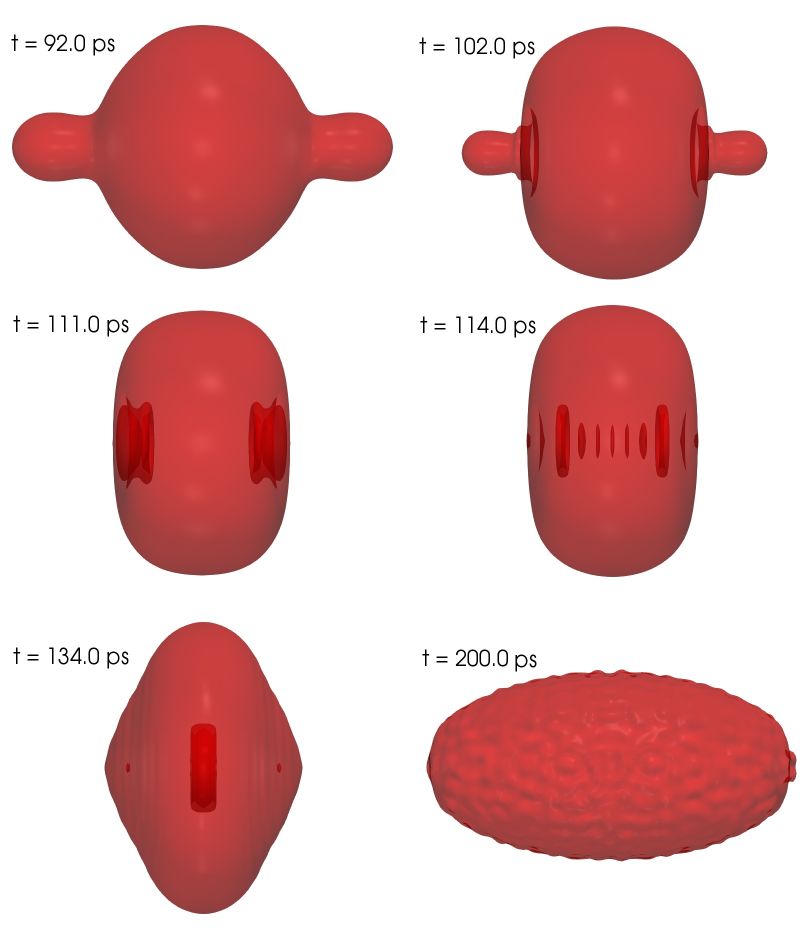}}
\caption{%
Sharp density surfaces, at the labeled times,
in the merging of two $^4$He$_{500}$ droplets.}
\label{fig2}
\end{figure}

Within DFT, the total energy of a $^4$He$_N$ droplet  at zero temperature
is written as a functional of an
effective wave function $\Psi( \mathbf{r},t)$
related to its atomic density by $\rho (\mathbf{r},t)= |\Psi( \mathbf{r},t)|^2$.
In this study, the number of helium atoms
in each of the initial droplets  is $N=500$, and the results we present have been obtained
using the 4He-DFT BCN-TLS computing package \cite{Pi17}.
The density functional used in this work is finite-range and includes
non-local effects \cite{Anc05}, since both aspects are needed to
describe quantitatively the response of the liquid at
the scale of the vortex core radius, which is of the order of
one Angstrom \cite{Gal14}. This functional
reproduces quite faithfully the density modulations around the vortex core \cite{Anc17}.
These modulations have been interpreted as a cloud of
bound virtual rotons embodied in the  phase of the vortex wave function, and
they may be converted into real ones following
vortex-antivortex annihilations, making vortex tangles
a potential source of non-thermal rotons \cite{Ame18}. Our simulations
provide a clean evidence of such conversion process.

We have first determined the equilibrium configuration of an isolated helium droplet
by solving the static DFT problem \cite{Anc17}.
These droplets are spherical, and their sharp density surfaces, defined
as the locus where the helium density equals half the bulk liquid value
$\rho_0 =0.0218$\,\AA$^{-3}$,
have a radius
$R=r_0 N^{1/3}$  with $r_0=2.22$ \AA{}.
The energy of the $^4$He$_{500}$ droplet is $-2474.6$ K,
and that of the $^4$He$_{1000}$ droplet is $-5400.3$ K. An energy of
$451.2$ K, mainly arising from the decrease of surface energy \cite{Bar06},
is thus available for the merging.

Next, we place the droplets at rest so that their surfaces
are about $6$\,\AA{} away from contact, and use this as the initial
configuration for the merging dynamics within the time-dependent DFT (TDDFT) approach.
This is in stark contrast with classical calculations of merging, since they have to start from
two sharp surfaces with a point of contact
and to assume that by some microscopic mechanism a tiny bridge appears joining the droplets \cite{Egg99};
and with BEC calculations of confined \cite{Sun08} or self-sustained droplets, where
some velocity has to be given to the droplets to trigger the collision \cite{Fer19}. In our case, it is the VdW
mutual attraction that provides such a microscopic mechanism.

The TDDFT equation \cite{Anc17,Pi17} has been solved up to
$t=600$\,ps with a time step $\Delta t=1$\,fs.
Droplet coalescence proceeds by the development of a tiny, low density bridge connecting the droplets in about $4.5$\,ps \cite{SM}.
Figure \ref{fig1} displays the
linear
$t$-dependence of the squared neck radius $R_{\textrm{neck}}^2$ with $6.31$ \AA$^2$/ps slope,
showing that $R_{\textrm{neck}} \propto t^{1/2}$. In spite of the different length and time scales,
the same behavior has been observed in experiments on low viscosity fluids \cite{Wu04}.
Calculations for classical droplets yielded \cite{Egg99,Duc03}
\begin{equation}
\frac{R_{\textrm{neck}}}{R} = \beta   \sqrt{\frac{t}{\tau}}
\end{equation}
with $\tau = (\rho R^3/\sigma)^{1/2}$,
where $\rho$ is the mass density
of the fluid, $\sigma$ its surface tension,
and
$R$ is the droplet's radius. 
The $\beta$ factor was calculated for inviscid droplets
yielding a value of 1.62 \cite{Duc03}, while measurements gave values
in the $1.03$--$1.29$ range \cite{Wu04}.
Using for  $\tau$ the liquid $^4$He value and
the calculated slope, we find $\beta = 0.93$.
We have analyzed the first few frames of Fig.~2 in Ref.~\onlinecite{Vic00},
also finding $R_{\textrm{neck}} \propto t^{1/2}$ but with a significantly higher prefactor, $\beta= 6.75$.
This discrepancy is attributed
to the different impact velocity of the droplets,
which is zero in our case and a few cm/s in the experiment \cite{Vic00}.

The droplets eventually merge, producing in the process two protrusions
symmetrically placed along the collision direction, as shown in Fig.~\ref{fig2}
for $t=92$\,ps.
The growth of the merging neck and the appearance of the
protrusions at the droplet surface displayed in
Fig.~\ref{fig2} are remarkably similar to those
found for the coalescence of $^4$He drops
of $0.20$--$0.25$\,cm radius, as
shown in Fig.~2 of Ref.~\onlinecite{Vic00},
with the only obvious differences being
the time and length scales involved.

As the protrusions shrink, pairs of quantized vortex-antivortex rings are
nucleated at the necks that connect the protrusions with the
merged droplet, with all vortex (antivortex) rings located around the left (right) neck.
Once formed, the rings of each pair move symmetrically towards the droplet center \cite{SM}.
Two such vortex-antivortex pairs can be
identified in Fig.~\ref{fig2} at $t = 111$\,ps. Additionally,
lenticular, sharp density indentations appear in the regions between
rings, likely shock waves produced by the sudden
collapse of the protrusions; these indentations are separated by a distance similar to that between He atoms
in the compressed liquid.

During their motion, the smaller radius
rings catch up to the larger radius rings and
pass through them, shrink and eventually disappear at
$t \simeq114$\,ps.
The larger radius vortex-antivortex rings last up to $t \simeq 139$\,ps, when
they collide against each other and annihilate, producing a roton burst
(as explained in the following). We may estimate the energy $\Delta E$ released in the annihilation event
by recalling that, in $^4$He, the energy of a vortex ring of radius $R_v$
is  $E_v = (2\pi ^2 \hbar ^2 /m)\rho _0 R_v [\ln(8R_v/a)-1.615]$ \cite{Rob71},
where
$a=0.7$\,\AA{} is the vortex core size.
In our case, the vortex ring radius just before annihilation
is $R_v\sim 8$\,\AA{},
which leads to $\Delta E= 2E_v
\sim  240$\,K. This corresponds to the emission of $\sim 24$ rotons \cite{Mor98}.

In the later stages of the time evolution, the droplet appears to be
in a highly excited state
characterized by short-wavelength nanoscopic surface waves,
as shown in Fig.~\ref{fig2} at $t=200$\,ps.
Similar surface patterns have been found in the decay
of multicharged vortices in trapped BEC \cite{Cid17}.
Superposed to this complex surface dynamics,
the merged droplet undergoes large amplitude oscillations, its shape
periodically shifting from oblate to prolate \cite{SM}.

\begin{figure}[hbt!]
\centerline{\includegraphics[width=1.0\linewidth,clip]{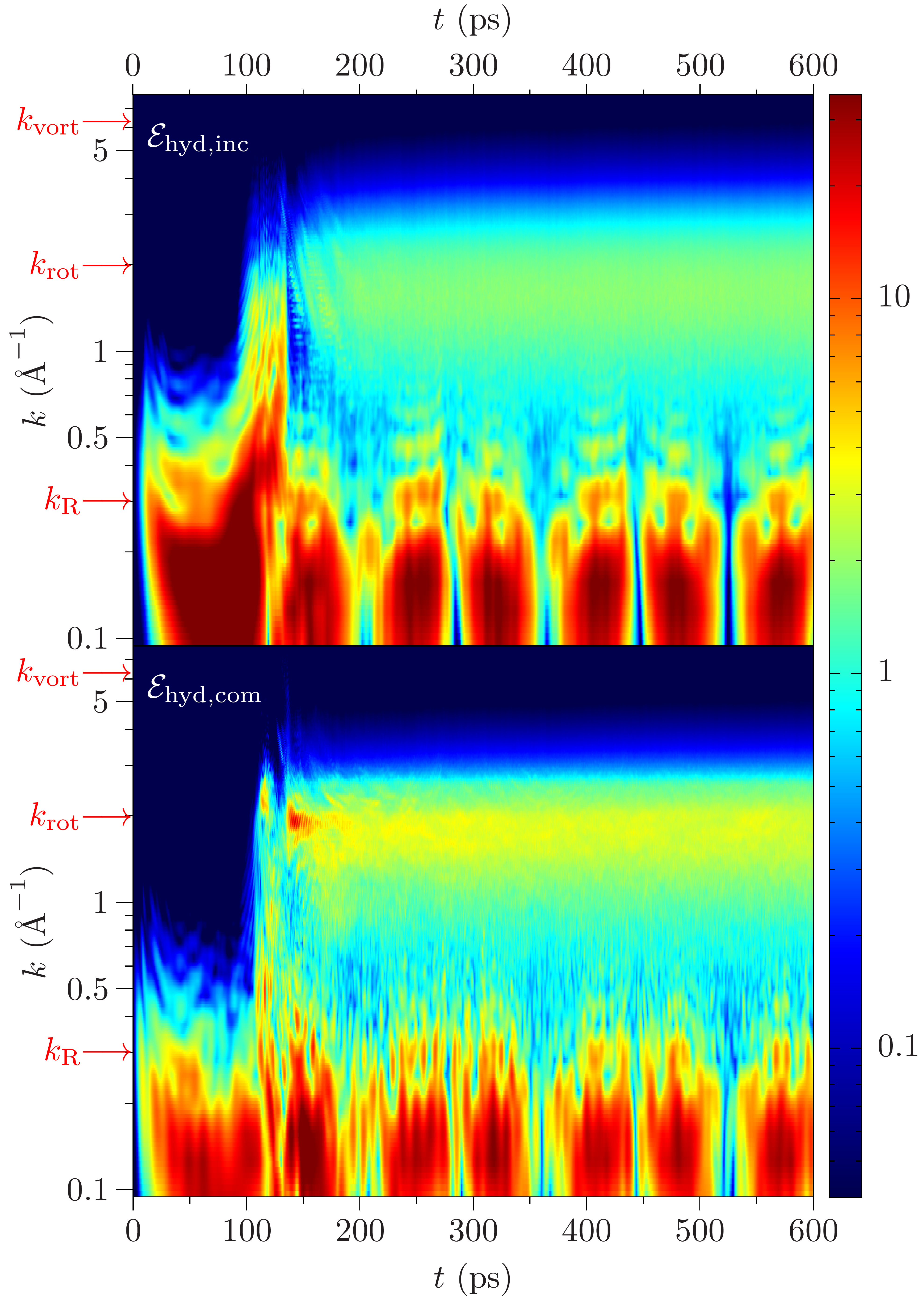}}
\caption{%
Energy spectrum $\mathcal{E}_{\textrm{hyd}}(k,t)$ (K\,\AA). Top: incompressible part; bottom: compressible part.}
\label{fig3}
\end{figure}

\begin{figure}[hbt!]
\centerline{\includegraphics[width=1.0\linewidth,clip]{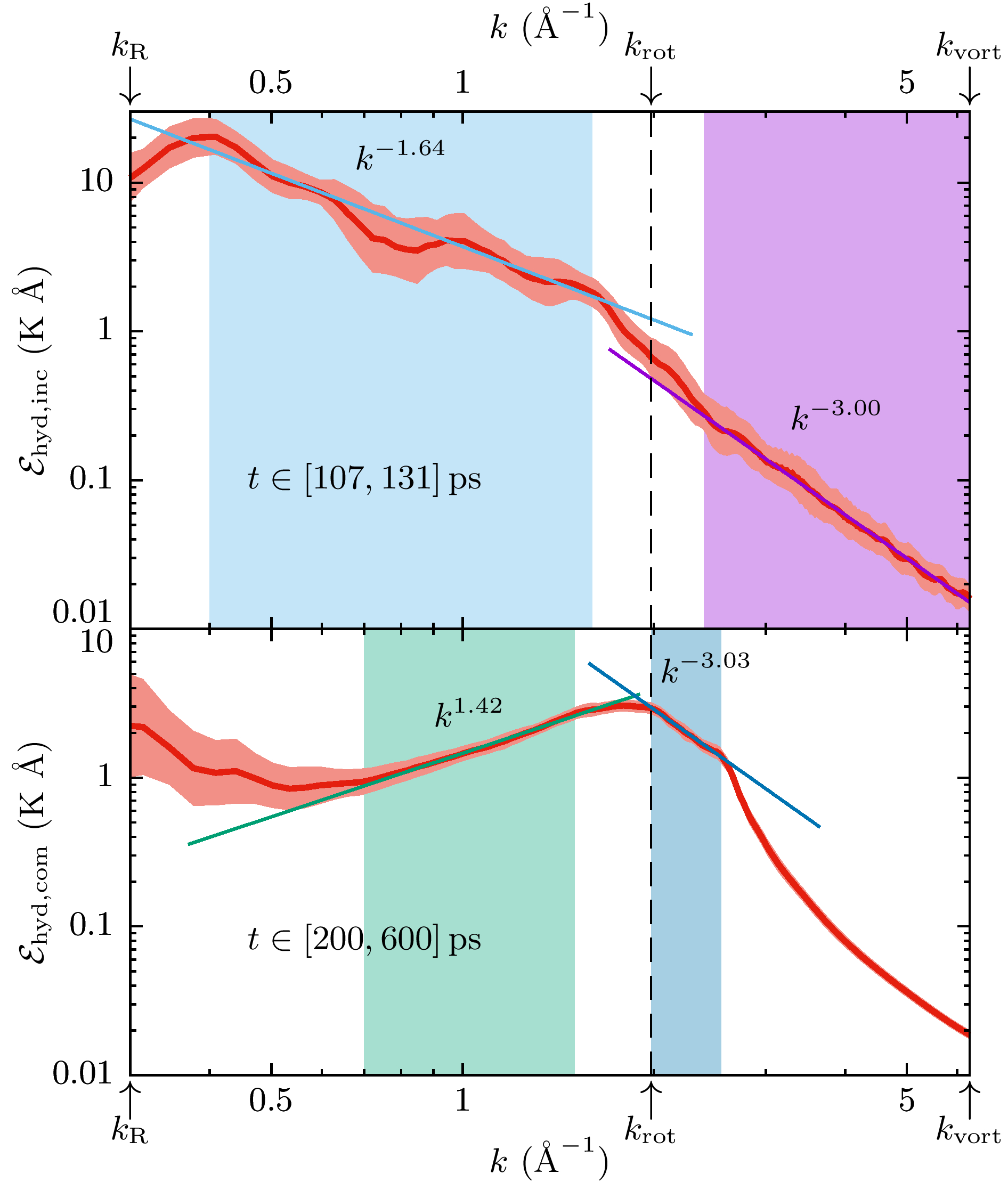}}
\caption{%
Power law analyses of time averages of the
incompressible (top panel) and compressible (bottom panel) energy spectra.
Each of the two panels displays, in log--log scales, the time-averaged
spectrum $\mathcal{E}(k)$
(red line) and the range within a standard deviation (light red) for every $k$,
as well as the corresponding power laws determined by weighted fits over
the shadowed $k$ intervals and indicated time intervals.
Top panel:
$k^{-1.64\pm0.05}$ for $k\in[0.4,1.6]\;\textrm{\r{A}}^{-1}$
(light blue)
and
$k^{-3.00\pm0.01}$ for $k\in[2.4,2\pi]\;\textrm{\r{A}}^{-1}$
(violet).
Bottom panel:
$k^{1.42\pm0.02}$ for $k\in[0.70,1.50]\;\textrm{\r{A}}^{-1}$
(green)
and
$k^{-3.03\pm0.05}$ for $k\in[1.98,2.55]\;\textrm{\r{A}}^{-1}$
(dark blue).}
\label{fig4}
\end{figure}

The presence of quantized vortices in the droplet during the merging,
their mutual interaction, and their
annihilation followed by the emission of rotons, is likely a
source of turbulence, which we address here
with the aid of a widely used tool in studies of classical and
quantum turbulence, i.e., the kinetic energy spectrum whose
dependence upon the wavenumber $k$ allows
to distinguish different regimes that
are relevant for characterizing turbulence \cite{Tsu09}.

Writing the effective wave function  as
$\Psi(\mathbf{r}, t)= \sqrt{\rho(\mathbf{r}, t)} \exp[\imath \,\mathcal{S}(\mathbf{r}, t) ]$,
the atom current density is
${\mathbf j}({\mathbf r},t ) = \rho({\mathbf r},t)  {\mathbf v}({\mathbf r},t)$ with
${\mathbf v}({\mathbf r},t) =   \hbar \, \nabla \mathcal{S}({\mathbf r},t)/m$.
Thus,  the kinetic energy of the superfluid can be written as
\begin{equation}
E_{\textrm{kin}}(t) =  \frac{\hbar^2}{2m}  \int \dd {\mathbf r} \, \left|\nabla \sqrt{\rho({\mathbf r},t)}\right|^2
+ \frac{m}{2} \int \dd {\mathbf r} \,  \frac{{\mathbf j}^2({\mathbf r},t)}{\rho({\mathbf r},t)}
\;.
\label{A1bis}
\end{equation}
The first term is the quantum pressure while the second is the usual hydrodynamic kinetic energy $E_{\textrm{hyd}}$.
Working in Fourier space, one can rewrite $E_{\textrm{hyd}}$ as
\begin{equation}
E_{\textrm{hyd}}(t) =  4 \pi \int_0^\infty  \dd{k}  \, \mathcal{E}_{\textrm{hyd}}(k,t) \; ,
\label{eq9}
\end{equation}
where the energy spectrum $\mathcal{E}_{\textrm{hyd}}(k,t)$ is the
spherical average in $\mathbf{k}$-space \cite{Bar01,Nor97,Bra12}
\begin{equation}
{\mathcal{E}}_{\textrm{hyd}}(k,t) = \frac{m}{2} (2 \pi)^3  \frac{k^2}{4\pi}
\int_{\Omega_k}  \dd\Omega_k
\left| \widetilde{\frac{ {\mathbf j}}{\sqrt{\rho}}}({\mathbf k},t)\right|^2
\;.
\label{eq10}
\end{equation}
We have decomposed it into a divergence-free (incompressible) part,
related to vorticity,
and a compressible part \cite{Nor97}, related to density waves,
and have analyzed them separately \cite{Ree12}.

Figure \ref{fig3} shows both additive components of $\mathcal{E}_{\textrm{hyd}}(k,t)$,
as well as three relevant $k$ values:
$k_{\textrm{vort}}= 2\pi/a = 6.3$\,\AA$^{-1}$, $a$ being the vortex core size $\sim 1$\,\AA{};
$k_{\textrm{R}}=2\pi/R= 0.3$\,\AA$^{-1}$, where $R$ is the radial dimension of the droplet, and
the roton wave-vector $k_{\textrm{rot}}=1.98$\,\AA$^{-1}$.
Both components periodically display bright regions in the small $k$ region
($k\lesssim k_{\mathrm{R}}$)
in phase with the oscillations of the shape of the merged droplet.
Interestingly, in the compressible part (bottom panel),
a bright peak appears  at $t\simeq140$\,ps corresponding to the roton burst
created by the annihilation of the large vortex-antivortex rings.
This peak spreads with time and is present in the rest of the simulation \cite{note1}.
The fainter spot at $t \simeq 115$\,ps originates from the  mentioned large amplitude density
waves that appear between the larger vortex and antivortex rings.

Between approximately $100$ and $140$\,ps, the relevant part of the incompressible energy spectrum
is dominated by the presence of  vortices and their decay (top panel of Fig.~\ref{fig3}).
The top panel of Fig.~\ref{fig4} corresponds to this regime.
It displays two powers laws, $k^{-3.00}$ and $k^{-1.64}$, which are
strikingly similar, respectively, to that arising from the presence of vortices \cite{Nor97,Bra12},
and to the Kolmogorov classical scaling, $k^{-5/3}$, which is known
to be also present in bulk superfluid turbulence \cite{Nor97,Sal10}.

At later times, the compressible part of the
energy spectrum is dominated by rotons and their effect on the
droplet surface (see below).
The bottom panel of Fig.~\ref{fig3} corresponds to this regime, whose scaling laws are shown in the bottom
panel of Fig.~\ref{fig4}.
It is the analogue of the weak-wave turbulence regime arising from acoustic radiation,
characterized by a $k^{-3/2}$ law \cite{Naz06,Tsu09},
although other scalings, such as $k^{-7/2}$, have also been found \cite{Ree12}.
In our case, the spectrum displays two powers laws with exponents $1.42$ and $-3.03$.
One might expect that the scalings associated to droplet surface turbulence are not those of two-dimensional
turbulence on a flat surface \cite{Naz06,Tsu09,Ree12}.  Moreover, since the wave energy spectrum
depends on the size and shape
of the droplet surface, one should not expect a universal behavior.

Inspection of the early stages of superfluid flow passing through the neck
connecting the two merging droplets shows
that the circulation lines have a tendency to bend outwards at the contact
region, hinting at the possible nucleation of
vortex-antivortex rings on the outer droplet surfaces, which cannot be observed
by the mere inspection of the droplet densities \cite{SM}.
To help identify the presence of vorticity
in these  low-density regions, we have calculated
the pseudo-vorticity $\nabla \times {\mathbf j}({\mathbf r},t )$ \cite{Vil16}.
Plots of  $|\nabla \times {\mathbf j}({\mathbf r},t )|$ iso-surfaces
allow to visualize regions of potentially non-zero vorticity \cite{Fre18}.

\begin{figure}[!tbp]
\centerline{\includegraphics[width=0.8\linewidth,clip]{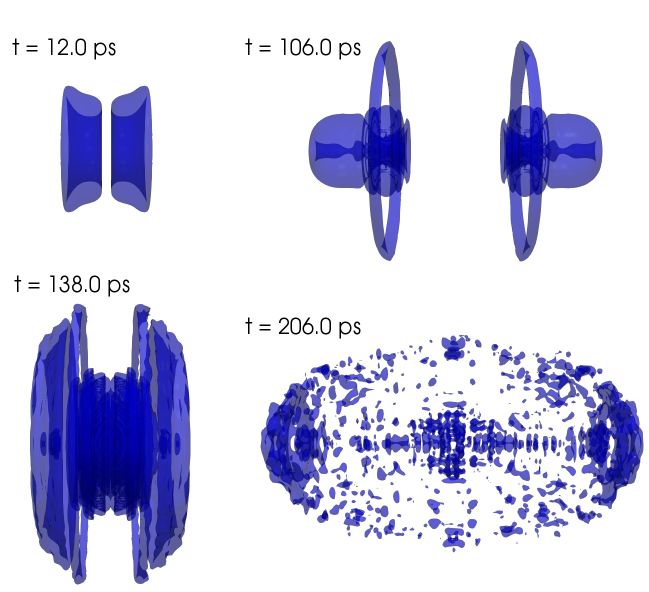}}
\caption{
Pseudo-vorticity  in the merging of two  $^4$He$_{500}$ droplets,  visualized by
$|\nabla \times {\mathbf j}({\mathbf r},t )|=10^{-3}$\,\AA$^{-3}$ps$^{-1}$  iso-surfaces
 at the labeled times.
}
\label{fig5}
\end{figure}

A video in the Supplemental Material \cite{SM} shows how pseudo-vorticity
spreads and slips on the outer surface of the merging droplets, and
Fig.~\ref{fig5} displays pseudo-vorticity
iso-surfaces for selected times.
We have  checked that,
between 12 and 90\,ps, the circulation around the more
intense of these tiny surface structures
is quantized, with
charges $\pm1$, hence signalling bona fide vortex/antivortex rings \cite{note}.
These vortices do not sink into the droplet but
remain on the surface (shells of parallel ring-like structures appearing in Fig.~\ref{fig5}).
Eventually, they decay and fragment  due to the
nanoscopic indentations appearing on the surface of the merged droplet
at $t \sim160$\,ps, when the surface changes its appearance
from smooth to rough \cite{SM}, thus causing the weak-wave turbulence discussed before.
Superfluid helium droplets, after capturing impurities \cite{Anc17} or being produced by
hydrodynamic instability of a liquid jet \cite{Gom14}, might experience a similar process.

Although most of the pseudo-vorticity appearing when $t>160$\,ps is localized on the droplet surface region,
lenticular patterns remain in its bulk. These patterns, which are remnants of the sharp density indentations
discussed before, can be seen, e.g., in Fig.~\ref{fig5} at $t=206$\,ps, where a train of such
faint structures is visible along the incident direction.

Finally, we would like to mention that
recent realizations of stable self-bound ultra-dilute quantum droplets, made of atoms of a binary mixture of BECs \cite{Cab18,Sem18},
allow to address droplet merging in a different superfluid environment \cite{Fer19}
and in a likely more controllable way, opening up the
possibility of studying wave turbulence on their surfaces. Self-bound BEC droplets made of dipolar gases displaying
a roton minimum \cite{Kla09} might also disclose non-thermal roton emission in these systems.

\begin{acknowledgments}
We thank Humphrey Maris, Luciano Reatto,
Jussi Eloranta, Alessio Recati and Albert
Gallem\'{\i}  for useful exchanges and discussions.
This work has been performed under Grant No  FIS2017-87801-P (AEI/FEDER, UE).
MB thanks the Universit\'e F\'ed\'erale Toulouse
Midi-Pyr\'en\'ees for financial support
throughout the ``Chaires d'Attractivit\'e 2014''  Programme IMDYNHE.
JME acknowledges financial support from the Engineering and
Physical Sciences Research Council, UK, under grant No.\ EP/P034616/1.
\end{acknowledgments}


\begin{thebibliography}{99}
\bibitem{note1}
Note that the roton burst is very intense because the droplets are identical and so are the colliding ring/antiring. 
If they were not, one would expect the intensity to decrease.
For large asymmetries, the rings might even cross without annihilating. Eventually, they should decay by interacting either 
with density modulations, as the small rings described here do, or with the droplet surface.
\bibitem{note}
Necessarily, at each chosen time, a segment of the closed circuit used to evaluate the circulation is in the low (but non-zero) density
region extending beyond the sharp density surface.
The contribution of this segment is needed to ascertain that the circulation is indeed quantized.
\end{thebibliography}
\end{document}